\documentclass[manuscript]{acmart}

\usepackage{makecell}

\AtBeginDocument{%
  \providecommand\BibTeX{{%
    \normalfont B\kern-0.5em{\scshape i\kern-0.25em b}\kern-0.8em\TeX}}}

\setcopyright{rightsretained}
\copyrightyear{2020}
\acmYear{2020}
\acmDOI{}

\acmConference[RecSys ’20]{CARS Workshop}{September 22–-26, 2020}{CARS Workshop, Online}
\acmVolume{}
\acmNumber{}
\acmArticle{}
\acmMonth{9}
\acmBooktitle{CARS 2020: Context-Aware Recommender Systems Workshop}

\begin{document}

\title{Session-based k-NNs with Semantic Suggestions for Next-item Prediction}

\author{Miroslav Rac}
\email{miroslav.rac@stuba.sk}
\affiliation{%
  \institution{Slovak University of Technology in Bratislava}
  \city{Bratislava}
  \country{Slovakia}
}

\author{Michal Kompan}
\email{name.surname@kinit.sk}
\affiliation{%
  \institution{Kempelen Institute of Intelligent Technologies}
  \city{Bratislava}
  \country{Slovakia}
}

\author{Maria Bielikova}
\email{name.surname@kinit.sk}
\affiliation{%
  \institution{Kempelen Institute of Intelligent Technologies}
  \city{Bratislava}
  \country{Slovakia}
}

\renewcommand{\shortauthors}{Rac, et al.}

\begin{abstract}
  One of the most critical problems in e-commerce domain is the information overload problem. Usually, an enormous number of products is offered to a user. The characteristics of this domain force researchers to opt for session-based recommendation methods, from which nearest-neighbors-based (SkNN) approaches have been shown to be competitive with and even outperform neural network-based models. Existing SkNN approaches, however, lack the ability to detect sudden interest changes at a micro-level, i.e., during an individual session; and to adapt their recommendations to these changes. In this paper, we propose a conceptual (cSkNN) model extension for the next-item prediction allowing better adaptation to the interest changes via the semantic-level properties. We use an NLP technique to parse salient concepts from the product titles to create linguistically based product generalizations that are used for change detection and a recommendation list post-filtering. We conducted experiments with two versions of our extension that differ in semantics derivation procedure while both showing an improvement over the existing SkNN method on a sparse fashion e-commerce dataset.
\end{abstract}





\maketitle

\section{Introduction}

Recommender systems can help users in fulfilling their intentions by suggesting the most relevant actions to follow. A wide variety of systems has been developed, while most of the approaches benefit from modeling of long-term user profiles whether of sequential or static nature. In many real-world applications, long-term user profiles are not available, and therefore, suggestions have to be made solely based on the observed actions during an ongoing session. Typically, session-based recommendation algorithms suggest a user the immediate next actions. The practical relevance of the problem leads to a number of proposals for algorithms mostly based on the analysis of item-level dependencies in interactional data.

When making recommendations based only on the click-stream data from an ongoing session, it is reasonable to take into account characteristics usually coming with the items. Users are likely to browse alternatives when they visit a website with the intention of purchasing a specific type of product. The alternative relations among products can be recognized by similar text descriptions. It has been shown that incorporating rich features into the recommendation process helps to deal with the sparsity and the item cold start, though it is not trivial \cite{hidasi2016parallel, tanjim2020attentive}.

\begin{figure}[h]
\centering
\includegraphics[width=0.6\textwidth]{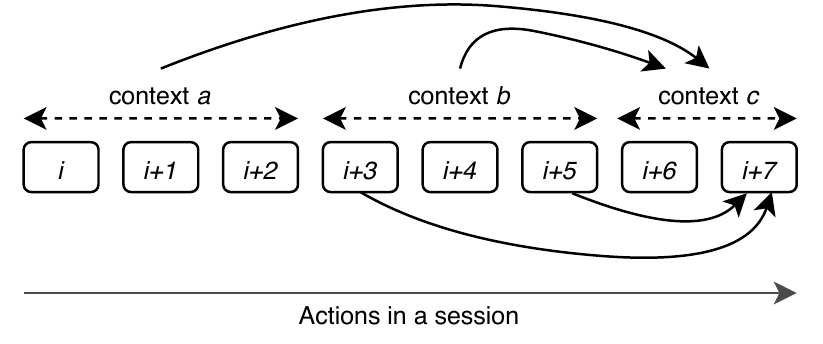}
\caption{Given that context is a relational property, an action can be related to its close neighborhood based on an arbitrarily devised precondition of coherence. As a result, a session is split into several sub-sequences of actions, within which context is seemingly constant and among which drifts can be easily detected. There are sequential dependencies expected, both on the level of actions and the level of sub-sequences.}
\label{fig:interactional-context}
\Description{figure description}
\end{figure}

Most of the existing session-based recommenders assume sessions to be associated with a single intention, however, it may change within an individual session multiple times. Detecting and adapting to these changes is one of the open challenges. Based on the idea of interactional context, a precondition can be devised that allows the delineation of coherent sub-sequences of semantically related actions (Fig.~\ref{fig:interactional-context}). For this purpose, we employ a technique for semantic analysis of natural language to extract high-level concepts from the item descriptions. We used it to help reduce unnecessary information that may introduce uncertainty into the intention change detection task. Within an individual session, users often browse items that complement the ones viewed earlier in the session. Hence the changes of semantics are intuitively expected to reflect patterns.

This paper is devoted to the hypothesis that boosting the score of semantically related items will improve the ranking of a recommendation list generated by a baseline method. The idea underlying our approach is to make a list of top-k recommendations and then re-rank the items in a way that items semantically related to the current intention will be moved to the top of the list while the order among them is maintained. 

The main contributions presented in this paper are as follows: 
\begin{itemize}
\item we extend a simple k-NN based recommender with a semantic factor that helps deliver recommendations on the user's current intention, the proposed cSkNN method improves Mean Reciprocal Rank (MRR) in all conducted experiments,
\item we compare effects of several linguistic-based generalization configurations on the recommended items ranking,
\item we evaluate two versions of semantic adaptation in the course of activity and compare their results to the potential given by the generalization configuration.
\end{itemize}

\section{Related works}

Traditional collaborative-filtering methods cannot be used as a session-based recommender because no past user behavior is available. A vital solution is the approach based on item co-occurrences in available click-stream data \cite{linden2003amazon}. Recommending items that are often clicked together has been proven to be effective in real-world applications and such nearest-neighbor-based algorithms of almost trivial nature are often used as competitive baselines. Session-based kNN method (SkNN), compared to Item-based kNN, considers an entire session when calculating the similarity, not only the last action. Several SkNN extensions providing sequential awareness are competitive with considerably more complex approaches, even those neural-network-based \cite{ludewig2018evaluation, jannach2017recurrent}.

Sequential recommenders are based on Markov model to utilize sequential dependencies. For example, Markov model makes predictions of latent topics derived from song tags that are then used to post-filter the initial ranking produced by a traditional kNN algorithm \cite{hariri2012context}. The simple Markov chains methods calculate transition probabilities depending only on the previous state. Higher orders may be used to model more complex dependencies, however, it dramatically increases the computational complexity. 

Recurrent neural networks (RNN) are better suited to model complex dependencies and are used in many areas. Gated recurrent units (GRU) can handle longer dependencies and achieve significant improvements over traditional methods \cite{hidasi2015session}.

Deep RNNs have the ability to capture sequential dependencies among a variety of entities and to involve rich features such as text descriptions or images \cite{hidasi2016parallel}, so naturally, the most of the research has been devoted to them in the past few years. Rich item representations have been utilized in Neural Attentive Session-based Recommendation model (NARM) to focus attention on  items similar to session purpose \cite{li2017neural}. There are also works that use such representations to divide a session. Mixture-channel purpose
routing networks (MCPRNs) model multiple intents in an individual session \cite{wang2019modeling} or semantic changes in attentive bi-GRU with song tags \cite{sachdeva2018attentive}. Attentive sequential
model of latent intent (ASLI) method uses categories to model the intent \cite{tanjim2020attentive}.

\section{Proposed approach}
The underlying idea of proposed cSkNN is to make a list of top-k recommendations and re-rank the items according to their semantics. In this section, we elaborate on our motivation and describe our method.

\subsection{Motivation}

Contextual information is an additional and important modeling dimension that helps to enhance the prediction accuracy. Existing approaches mostly focus on a \textit{representational view} of the context. It is a descriptor of an environment in the form of stable information that can be delineated in advance and then simply observed before activity or during its initialization, e.g., weather, time, or a device’s sensory inputs. However, we lack the richness of such relevant contextual information in e-commerce domain so we take the position of an alternative \textit{interactional view} that differs from the former fundamentally. Instead of context being considered as a bit of information, it is treated as a relational property that is actively produced and maintained in the course of activity \cite{dourish2004we}. A session-based recommender is a kind of a protagonist of such a phenomenological position since a session can be seen as context in the sense of a property relating actions together.

However, context is perpetually being altered so when context is represented with a session as a whole, it may contain misleading information that is not up-to-date after several actions. Since a system cannot monitor all factors inducing changes in context, a temporal factor is often used as an acceptably effective proxy for these hidden factors. Researchers proposed a variety of functions that decay weight of actions in time or attention mechanisms to assign significance to the most recent actions; and/or to actions related to them \cite{li2017neural, garg2019sequence, guo2018adjustable, campos2014time}. The drawbacks of using temporal information as a solitary indication of a change are that there is no indication of a preference shift direction, and also that widely-used decaying functions lack the ability of adaptation to sudden changes within an individual session.

Labeling each action with a high-level concept conveys a reliable coherence precondition that is needed for the drift detection. Generalization of actions that enforces high-level similarities and repress low-level differences (i.e., noise) will indicate the exact boundaries of sub-sequences as depicted in Figure~\ref{fig:interactional-context}. Representations learned by the distributional analysis of the interactional data inherently give rise to generalizations. However, this approach is less useful if a sufficient amount of user-item interactions is not available and also introduces bias from user interactions (as such information is already available on the level of items).

Therefore, we opt to exploit a concise nature of item titles in e-commerce catalogs, from which high-level concepts can be obtained via semantic analysis. Despite difficulties emerging from the akin nature of the text, dependency parsing enables a higher control over the process of semantic generalization compared to the analysis based on the distributional hypothesis or a popular bag-of-words approach.

Intuitively, transitions among coherent sub-sequences follow patterns. For example, users tend to see complementary accessories (e.g., belt) along with items that exhibit the origin purchase intention (e.g., trousers), and also users might browse items that together constitute a whole outfit. A sequential pattern of concepts “trousers” → “belt” would have higher support and confidence than sequences of particular items that correspond to those concepts. Since all such patterns on the item-level are aggregated to a single generalized pattern on the concept-level. This helps to tackle the sparsity and possibly the cold start problem.

As SkNN methods are already competitive with the more sophisticated approaches, we devote more effort to their development. We evaluate a version with a simple extension (S-SkNN) that makes the method aware of sequential dependencies \cite{ludewig2018evaluation}. The S-SkNN method is used for predicting both items and semantics used to post-filter a set of recommended items.

\subsection{Semantic clusters as a coherence precondition}

We make a division of items regarded as having particular shared characteristics expressed by natural language into mutually exclusive \textit{semantic clusters}. These will satisfy the coherence precondition that allows splitting a session into several coherent sub-sequences among which is easy to detect interest drifts.

\begin{figure}[h]
\centering
\includegraphics[width=0.5\textwidth]{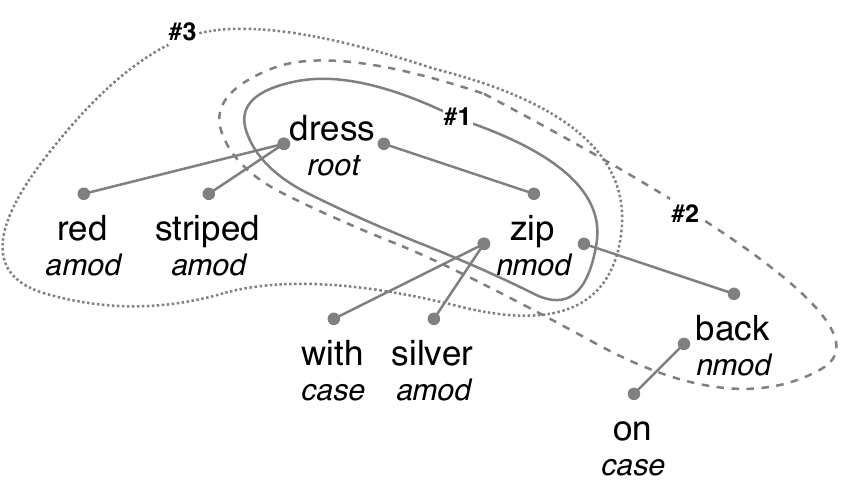}
\caption{Visualisation of a dependency tree with for example input title “Red striped dress with silver zip on back”}
\label{fig:configurations}
\Description{figure description}
\end{figure}

Inferring intentions from items is more difficult than inferring them from categories \cite{tanjim2020attentive} because they ease interaction sparsity issues and also carry explicit information about item meaning for a user. In e-commerce, human-made categories have a role of such semantic clusters, however, they are mostly widely defined and contain hundreds or thousands of items (e.g., \textit{summer dress}). The larger clusters the more items of a higher variability they contain, hence it implies a sort of hierarchical organization in which a variety of items are aggregated into a single cluster. This allows bound actions to their neighbors into semantic windows as depicted in Figure~\ref{fig:configurations}. Since the length of windows is directly affected by the size of clusters, we need to strike a balance between their size and the level of abstraction that allows both to detect drifts of interest and effective semantic filtering of recommendations. The size of semantic clusters has to be relatively smaller than the usual category size, however, not too small, because it would reduce the effectiveness of sparsity reduction and make the semantics prediction harder.

Besides the size also the constitution of clusters is important because by merely enlarging clusters we can cause items that do not satisfy the convenient coherence condition to merge into a single cluster. Generalization based on natural language processing can help to overcome this issue. Dependency parsing is the task of recognizing a syntactic structure in a sentence, hence it allows a controlled extraction of words that carry a semantically essential part of item characteristics.

The main goal of the product title is to attract the attention of potentially interested users, and therefore, titles are written rather concisely and mostly contain words describing the main characteristics of the product along with several adjuncts. Although we can exploit such nature, it also often leads to syntactically or grammatically incorrect constructions that considerably complicate the semantic analysis. We identified several issues and inconveniences, such as multiword and multi-language expressions or inflections of words, however, in our experiments we put the only reasonable effort in resolving the most critical ones. We believe that more effort would improve the end results, so we will discuss some of the identified issues briefly in Section \ref{sec:discussion}.

\begin{table}[h]
  \caption{Definitions of generalization configurations and example inputs for item title {\itshape ``Red striped dress with silver zip on back''}}
  \label{tab:confs}
  \begin{tabular}{cll}
    \toprule
    Conf.&Extraction rule&Example\\
    \midrule
    \#1 & root + first degree nmod & dress zip\\
    \#2 & root + second degree nmod & dress zip back\\
    \#3 & root amods + root + first degree nmod & red stripes dress zip\\
    \#4 & all words & red stripes dress silver zip back \\
  \bottomrule
\end{tabular}
\end{table}

The idea is to extract salient words rather linguistically than statistically as via TF-IDF method. Dependency parsing help to keep only highly distinctive descriptors and to reduce low-level noise, e.g., infrequent colors, causing the items to be merged in an undesirable way. The key part of parsing is the nominal phrase root extraction since it has a function to express a high-level concept which is in collocation with modifiers that add descriptive information to it \cite{de2014universal}. We focus on nominal dependents (\textit{nmod}) because they function as a non-core argument that brings about a relatively dense packaging of referential information, e.g., \textit{dress with zip on back}, which results in an extreme reliance on implicit meaning providing a utility for our task. Another important function is an adjectival modifier (\textit{amod}) which modifies the meaning of a nominal, e.g., \textit{striped dress}. Usually, it describes qualities of items such as material, color, or other distinctive appearance or functions.

Table \ref{tab:confs} describes four different rules of the generalization process from the standpoint of dependency parsing which control what words are extracted and used to create high-level concepts for items. We call them configurations in this paper from now on.

We need to create semantic clusters to use their \textit{IDs} during recommendation using a k-NN method. After words are normalized, we use \textit{word2vec} model trained on all item titles to create semantic relations among similar concepts which is useful in the process of generalization during the clustering process. Each item is then represented by an average of vectors of extracted words concatenated with an item category vector. This allows to distinguish among likewise named concepts, e.g., t-shirt in women and in men category, and the nature of the category tree also allowed to capture relations based on the parts of an outfit, e.g., upper or lower body.

We used a density-based clustering method DBSCAN to group items together with their surroundings given by an $\epsilon$ parameter. 
The radius given by $\epsilon$ directly influences the number of clusters and their sizes, therefore, it is another factor that strongly influences the generalization process alongside the word extraction rules.
We used different $\epsilon$ values according to the distributions of nearest neighbors' distances for each configuration.
As the result, we have 29 different mappings and for each of them, we have a function $cmap:I \rightarrow C$, where $I$ is a set of all items and $C$ is a set of all semantic clusters for a particular semantic cluster configuration.

\subsection{Recommendation methodology}

To make recommendations, we use S-SkNN approach with an extension providing sequential awareness: given the current session $s$, its neighbors $N_s$ and a pairwise similarity function, the recommendation score for an item $i \in I$ is defined as:
$$score_{S-SkNN}(i,s) = \sum_{n \in N_s} sim(s,n) \cdot w_n(s) \cdot 1_n(i)$$
where the similarity function is a dot product of binary-encoded sessions,
the indicator function $1_n(i)$ returns $1$ if session $n$ contains item $i$ and $0$ otherwise.
If an item $s_x$ is the most recent item of the current session and it also appears in the neighbor session $n$, then the weight is defined as $w_n(s)=x/|s|$, where the index $x$ indicates the position of $s_x$ within the session.
Hence, the weight increase depends on the position of an item that appears in both sessions, while the more recent position in the current session gains a higher weight. More information can be found in \cite{ludewig2018evaluation}.

To re-rank recommendations and favor semantically related items, we extend the scoring function by the semantic factor:
$$score_{cSkNN}(i,s) =score_{S-SkNN}(i,s) + c_s(i)$$ 
where $c_s(i)$ returns $1$ if an item $i$ is relevant to semantics given by $s$ and $0$ otherwise.
Since the S-SkNN score is normalized into a range of $<0,1>$, the $c_s(i)$ factor assures the semantically relevant items to appear at the beginning of the recommendation list.
We evaluate two different implementations of the semantic factor functions which differ in the complexity of generating a semantic cluster candidate.

\subsection{Semantic candidates by the last action \label{sec:last-semantics}}

Based on the assumption that semantics does not change along with several actions, we evaluate a version of a model that favors items semantically related to the last known action.
As a downside, the semantic factor function suggests wrong semantics at each beginning of a coherent sub-sequence. On the other hand, it is able to adapt immediately and there is some assurance that it will be correct for several following actions while it depends on the quality of semantic clusters.
In our interaction data, the average length of semantically coherent sub-sequences is roughly from $1.5$ to $3.8$, depending on the clusters' configuration.
In this version, $c_s(i)$ returns $1$ if $cmap(i)$ equals to $cmap(s_{|s|})$, where $s_{|s|}$ is the last known action in the session $s$.

\subsection{Prediction of semantic candidates \label{sec:predicted-semantics}}
Semantics is derived not only from the last known action but rather from the entire session. 
During the training phase, another S-SkNN instance is trained on the train sessions mapped on semantic clusters from a particular $C$ configuration using the $cmap(i)$ function. The instance can be then used for the prediction of the next immediate semantic cluster.
To make an effect with semantic filtering, we have to rely on the top-1 prediction according to semantic-level dependencies, which is used to improve the score of semantically related items in the items’ recommendation list. 
Therefore, $c_s(i)$ returns $1$ if $cmap(i)$ equals to a semantic cluster given by top-1 prediction using $score_{S-SkNN}(c, s')$, where $c \in C$ and $s'$ is the current session mapped onto the configuration $C$.

\section{Experiments}

\subsection{Dataset}
We conducted our experiments on a fashion e-commerce dataset that contains approx. 572k actions in 100k sessions in a period of 14 days.
The number of users is unknown since sessions are anonymous, therefore, only short-term dependencies within an individual session can be analyzed.
The dataset is very sparse, the density calculated by session-item interactions is 0.011\%.
The dataset contains approx. 155k products organized into 366 categories of shallow hierarchical structure with the top-level categories dividing items by the gender and age diversity dimensions, i.e., women, men, girls, and boys. The lower 3 levels divide items by the parts of the outfit, e.g., shirts, shoes, or accessories. Although our semantic analysis is based on all of the items, only 39k of them are used in the interaction data. As the texts are in the Czech language, we used UDPipe tool with Czech model \cite{udpipe:2017} for dependency parsing.

\subsection{Evaluation protocol}
We use an evaluation scheme in which the task is to predict the immediate next item given the first $n$ actions of a session.
For each session, we iteratively increment $n$ and make predictions. This simulates user activity.

Each time a prediction is made, we measure two popular top-k metrics in recommender systems, namely: hit rate (HR@k) and Mean Reciprocal Rank (MRR@k) for a set of \textit{k} values.
HR measures the proportion of cases having the true item amongst the top-k items, which is a suitable metric for certain scenarios where absolute order of recommendations does not matter.
MRR is important in cases where the order of recommendations matters, which is our case since the main purpose of our task is to re-rank items and to prefer items by the semantics of the current user intent. Both the reciprocal rank and hit rate are set to zero if the rank is above \textit{k}.
The dataset is divided into random train and test subsets on the level of sessions in a ratio of 80:20, each configuration is evaluated on the same train-test sample. 

For the comparison, the following types of recommenders are reported:
\begin{itemize}
    \item {\bfseries baseline}: a vanilla S-SkNN model without the semantic factor with a random sampling of 1000 neighboring sessions to speed up the prediction. Therefore, we cached the scores outputs given by $score_{S-SkNN}(i,s)$ and used them in the evaluation of all semantic configurations and methodology versions.
    \item {\bfseries last semantics}: a cSkNN model using the last seen semantics as described in Section \ref{sec:last-semantics}.
    \item {\bfseries predicted semantics}: a cSkNN model using the semantics predicted by the entire session as described in Section~\ref{sec:predicted-semantics}. Different S-SkNN instance is trained for each semantic clusters configuration separately. Only 25\% of the most recent actions in the train set is used to train the semantics prediction since more of the data did not lead to significant improvement because generalization produces a large number of identical sequences.
    \item {\bfseries potential}: to show the maximum theoretical improvement, we employ a version of the $c_s(i)$ function whereby it uses the true next immediate semantic cluster via a leak from the future.
\end{itemize}
For all schemes with the semantic factor, we re-rank top-50 recommendations generated by the baseline model.

\subsection{Results}

The evaluation showed that semantic post-filtering increases the performance even though the semantic prediction accuracy is relatively low. One of the most important takeaways is that the generalization via NLP has a higher potential for improving recommendations compared to the popular bag-of-words approach which is manifested by configuration \#4. Table~\ref{tab:conf-all-k} shows that MRR is improved in higher rates compared to HR, i.e., the method does not find more of the relevant items but ranks them better.

\begin{figure}[t]
  \centering
  \includegraphics[width=\linewidth]{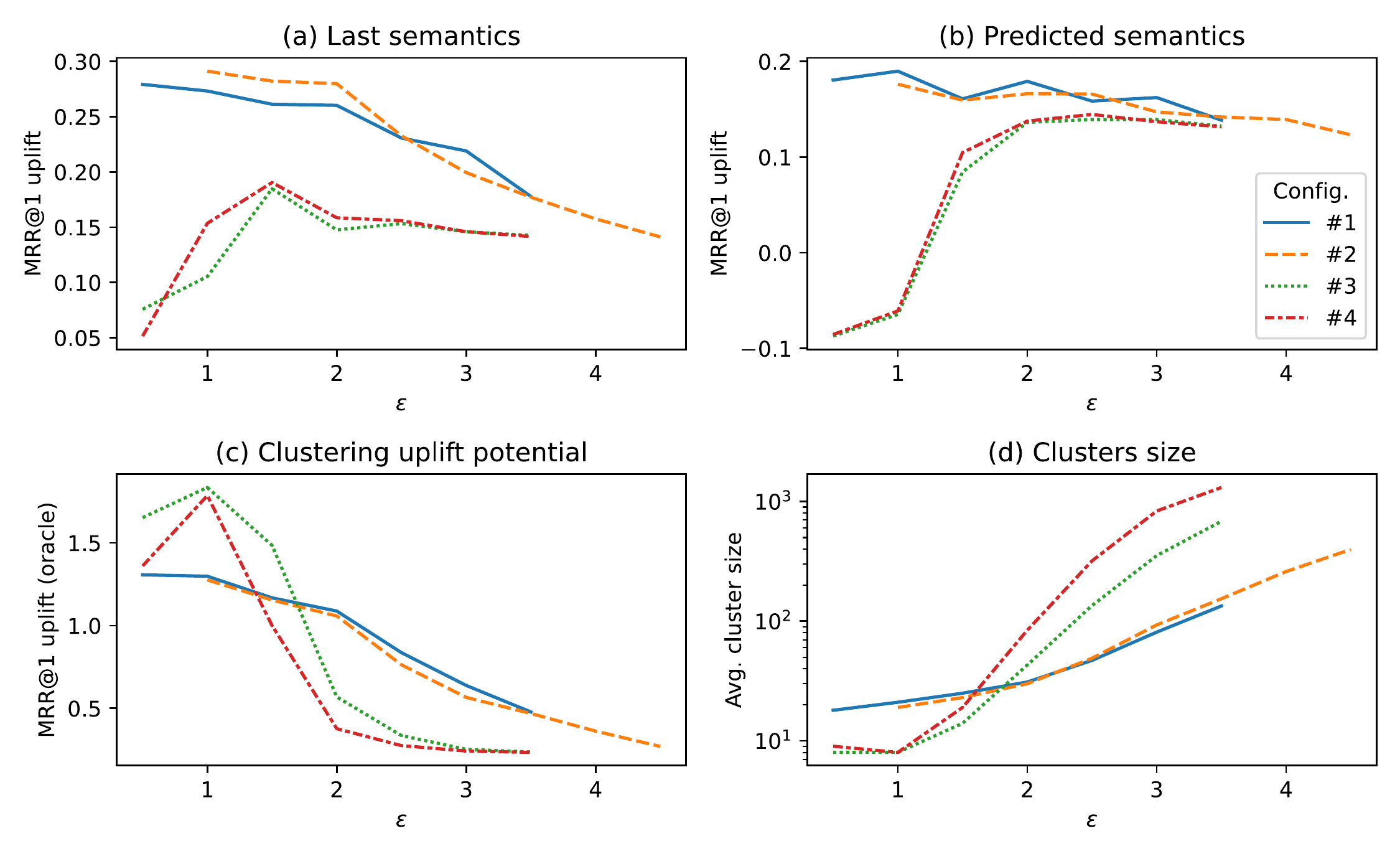}
  \caption{Comparison of generalization effects produced by the combination of different NLP parsing configurations and $\epsilon$ parameter of DBSCAN clustering. Line charts show different values given by the growth of $\epsilon$ for each generalization configuration. Configurations using high-level concepts (\#1 and \#2) have stronger linear correlation of $\epsilon$ with metric uplift in all semantic recommendation schemes.}
  \label{fig:conf-effects}
  \Description{}
\end{figure}

Figure~\ref{fig:conf-effects} shows noticeable differences between configurations which convey high-level concepts given by only \textit{nmods} (\#1, \#2) and the other configurations that use also \textit{amods} (\#3, \#4).
Configurations \#1 and \#2 provide consistently higher MRR@1 uplift than \#3 and \#4. 
We believe it is because the former omit low-level features (qualities such as colors etc.) that can be derived from the item-level dependencies but cause undesirable merging on the level of semantics.
Just as is the case of \#3, if conceptually unrelated items are merged because of their identical appearance quality aspects, clusters must be large enough to create coherent sub-sequences in the browsing behavior.
Although large clusters are easier to predict since their quantity declines with the growing size, the trade-off is unfavorable in this setup because the larger clusters the lower potential for the performance uplift.
Configurations \#3 and \#4 display MRR growth (Fig.~\ref{fig:conf-effects}a,b) stopping at the point where clusters become too large and their potential consequently starts to decrease significantly (Fig.~\ref{fig:conf-effects}c).

The last semantics scheme improves performance in all cases. It benefits from the length of intrasession coherence, however, results show a strong negative correlation between the coherence and the performance uplift for \#1 and \#2; and weak positive correlation for \#3 and \#4.
This means that conceptual coherence is inherently given by user behavior and increasing it artificially by employing larger semantic clusters is not inevitably beneficial as can be also seen in Table~\ref{tab:conf-results-overall}.

On the other hand, we noticed many cases of concepts alternating in sequences, e.g, $[a,a,b,a,b,a]$, what incapacitates advantages of the last semantics scheme, therefore, improving coherence by also incorporating information about interactions among the concepts prior to the clustering process could help.
In this scheme, the $c_s(i)$ function accuracy calculated as HR@1 is only $\sim$35\% for the best clustering configurations.
Despite flaws that arise from dynamics, this scheme ensures providing the correct semantic cluster within the coherent window regardless of the cluster's size or popularity which is not the case in the predicted semantics scheme.

Also, the predicted semantics scheme improves performance though in smaller rates. The \mbox{top-1} cluster prediction accuracy is roughly the same as in the last semantics scheme with small fluctuations around $\pm$1-2\%. Nevertheless metrics uplift is in most cases lower, because large clusters that provide lower utility are often predicted since the SkNN models tend to suffer from the popularity bias. Moreover, it might have trouble predicting the same cluster several times in case of a longer coherent semantic window. Therefore, this scheme is less suitable for the next-item prediction but it could be useful in different use case scenarios as discussed in Section~\ref{sec:discussion}.

In our settings, the semantic post-filtering has the potential to double the performance (Fig.~\ref{fig:conf-effects}c). However, the best performing mappings utilized only up to $\sim$20\% of the potential as shown in Table~\ref{tab:conf-results-overall}. Hence, there is still the opportunity for improvement of the semantic factor function $c_s(i)$.

\begin{table}[t]
\caption{The results for the best mapping in each configuration and its characteristics in the means of average cluster size and average length of coherent sub-sequences. The table shows achieved performance uplifts and compares them to the maximum theoretical potential given by the dedicated scheme. Also the rate of the potential utilization is shown. Note that at k=1, MRR and HR are equal.}
\label{tab:conf-results-overall}
\begin{tabular}{lccclcc}

\toprule
    \multicolumn{1}{l}{}	&	Conf.    & \thead{Avg. \\ coherence} & \thead{Avg. \\ cluster size} & \thead{MRR@1 \\ HR@1}    & \thead{Potential \\ MRR@1} & \thead{Utilized \\ potential} \\
\midrule
 baseline &&          1.15 &                1 & 3.20\% &                - &               -  \\
 \hline
 last      
       & \#1 &          1.77 &               18 & 4.09\% (+27.93\%) &   7.38\% &             21.37\%  \\
       & \#2 &          1.78 &               19 & \textbf{4.13\% (+29.12\%)} &   7.28\% &             22.82\%  \\
       & \#3 &          1.63 &               14 & 3.79\% (+18.48\%) &   7.95\% &             12.45\%  \\
       & \#4 &          2.07 &               19 & 3.81\% (+19.05\%) &   6.40\% &             19.04\%  \\
 \hline
 predicted 
 	  &      \#1 &          1.78 &               21 & \textbf{3.81\% (+18.98\%)} &   7.35\% &             14.62\%  \\
      & \#2 &          1.78 &               19 & 3.76\% (+17.62\%) &   7.28\% &             13.81\%  \\
      & \#3 &          3.54 &              135 & 3.64\% (+13.93\%) &   4.27\% &             41.53\%  \\
      & \#4 &          3.62 &              319 & 3.66\% (+14.46\%) &   4.08\% &             52.66\%  \\
\bottomrule
\end{tabular}
\end{table}

\begin{table}[t]
\caption{The results for the best mapping in each configuration. The table shows achieved relative performance uplifts compared to the baseline. The best results are highlighted for each metric and scheme. Note that at k=1, MRR and HR are equal.}
\label{tab:conf-all-k}
\begin{tabular}{lcrrrrr}
\toprule
& Conf. & MRR@1 & MRR@10 & MRR@20 & HR@10 & HR@20 \\
\midrule
baseline & &    3.20\% &                  5.53\% &                  5.62\% &                10.84\% &                12.16\% \\
\hline
last &  \#1 &       +27.93\% &                 +16.56\% &                 +15.86\% &                 +4.27\% &                 +1.14\% \\
    &  \#2 &        \textbf{+29.12\%} &                 \textbf{+17.19\%} &                 \textbf{+16.47\%} &                 \textbf{+4.39\%} &                 +1.57\% \\
    & \#3 &         +18.48\% &                  +9.30\% &                  +8.92\% &                 +3.59\% &                 \textbf{+1.63\%} \\
    & \#4 &         +19.05\% &                 +10.28\% &                  +9.85\% &                 +3.69\% &                 \textbf{+1.63\%} \\
\hline
predicted
        &  \#1    &   \textbf{+18.98\%} &                 \textbf{+11.29\%} &                 \textbf{+10.71\%} &                 +3.56\% &                 +0.84\% \\
        &  \#2    &   +17.62\% &                 +10.31\% &                  +9.79\% &                 +3.18\% &                 +1.48\% \\
        &  \#3    &   +13.93\% &                  +8.86\% &                  +8.50\% &                 +3.32\% &                 \textbf{+1.57\%} \\
        &  \#4    &   +14.46\% &                  +9.36\% &                  +8.94\% &                \textbf{ +3.59\%} &                 +1.54\% \\
\bottomrule
\end{tabular}
\end{table}

\section{Discussion and conclusions \label{sec:discussion}}

In this paper, we presented a proof-of-concept method that allows better adaptation of the next-item recommendations to the user interest via the semantics derived from product titles. The main contribution is the finding that the item generalization based on semantic analysis appears to provide a high potential for improving the recommendations. We extended a popular session-based k-NN recommendation model to prototype our idea, though it is not limited to any kind of model.

We employed the dependency parsing technique to identify head-dependent relations in a noun phrase describing an item and to extract only the most salient words that convey high-level concepts. The results of our experiments conducted on a fashion e-commerce dataset showed that omitting adjunct words describing low-level quality aspects of items, such as colors, leads to better items ranking in recommendation list when using text features to improve model performance.

The experiments showed that NLP based semantic analysis has significant potential in improving recommendations, from which the large part is not utilized yet. In the future, we plan to design the interest change detection as a binary task that can use behavioral data or other events, such as category view or add to cart, to switch between the last and the predicted semantics schemes. Furthermore, we plan implementation for the neural-network-based method, for which a latent representation can also be envisioned that would allow us to handle headwords and modifiers in a more flexible way.

We believe that the quality of semantic clusters and their positive effects on the performance can be improved even further. We identified several complications emerging from the nature of the text that had a bad influence on dependency parsing. The crucial part is to identify the root headword correctly, which is problematic when the root is a multiword expression, such as \textit{skinny jeans}. Also, the correct root headword may be misleading in some cases, e.g., \textit{pair of socks} would be inconveniently matched with \textit{pair of gloves} because of the common root \textit{pair}. Sometimes redundant words are used, e.g., \textit{woman dress} in a woman category is no different from products that do not contain the word \textit{woman}. The semantic analysis must be able to deal with copywriters' different styles, e.g., \textit{red dress} and \textit{dress in red color} have a different constitution but essentially the same meaning. We did not put effort into resolving these so a more comprehensive semantic analysis might provide a higher improvement.

Although we evaluated the proposed method on the next-item prediction task, it can be also used in different scenarios, e.g., as a shopping cart page recommendation strategy. After a user adds an item to the cart, top-n next-concept prediction can be used to split recommended items into \textit{n} recommendation lists for each concept separately, while we can rely upon that the sequential dependencies from the preceding browsing behavior would be also considered.

\begin{acks}
This work was partially supported by the Slovak Research and Development Agency under the contract No. APVV-15-0508 and the Scientific Grant Agency of the Slovak Republic, grant No. VG 1/0667/18. The authors would like to thank for financial contribution from the STU Grant scheme for Support of Young Researchers.
\end{acks}

\bibliographystyle{ACM-Reference-Format}
\bibliography{acmart}


\end{document}